\documentclass[journal=jacsat,manuscript=article]{achemso}
\usepackage{chemformula} 
\usepackage[T1]{fontenc} 
\usepackage{amsmath}

\author{Elijah Flenner}
\affiliation{Department of Chemistry, Colorado State University, Fort Collins Colorado}
\email{flennere@gmail.com}
\author{Grzegorz Szamel}
\affiliation{Department of Chemistry, Colorado State University, Fort Collins Colorado}
\email{grzegorz.szamel@colostate.edu}

\title[Defects]
  {Defects, Sound Damping, and the Boson Peak in Amorphous Solids}

\abbreviations{IR,NMR,UV}
\keywords{American Chemical Society, \LaTeX}

\begin{document}

%
%
%
%
%

\begin{abstract}
Two nearly universal and anomalous properties of glasses, the peak in the specific heat and plateau of the thermal conductivity, occur around the same 
temperature. This coincidence suggests that the two phenomena are related. Both effects can be rationalized by
assuming Rayleigh scaling of sound attenuation and this scaling leads one to consider scattering from defects. Identifying defects in glasses,
which are inherently disordered, is a long-standing problem that was approached in several ways. We examine candidates for defects in glasses that 
represent areas of strong sound damping. We show that some defects are associated with quasi-localized excitations, 
which may be associated with modes in excess of the Debye theory.  We also examine generalized Debye relations, which relate sound damping and
the speed of sound to excess modes. We derive a generalized Debye relation that does not resort to an approximation used by previous authors.
We find that our relation and the relation given by previous authors are almost identical at small frequencies and also reproduce the independently
determined density of states. 
However, the different generalized Debye relations do not agree around the boson peak. While generalized Debye relations accurately predict the 
boson peak in two-dimensional glasses, they under estimate the boson peak in three-dimensional glasses.  
\end{abstract}

\section{Introduction}
Understanding the origin of the excess density of states above the Debye theory in glasses is essential to explaining their anomalous low-temperature
properties \cite{Ramos2023,Zeller1971,Buchenau1984,Schroeder2004}. 
One such property is the boson peak, \textit{i.e} the peak in the reduced density of states, $g(\omega)/\omega^{d-1}$, where $d$ is the dimension.
The boson peak is related to the peak in the heat capacity around 10K. 
The nature of the vibrational modes around the boson peak and the source of these modes remain important questions \cite{Moriel2024}. Another anomalous
property is the plateau of the thermal conductivity that occurs around the same temperature as the peak in the heat capacity, which suggests
a connection between the two observations. 

The specific heat and thermal conductivity can be reproduced by fitting a few parameters of the soft potential model. 
At the heart of 
the soft potential model is the existence of quasi-localized excitations that leads to $\omega^4$ scaling of the 
density of excess modes\cite{Ramos2023,Karpov1982,Buchenau1992,Ramos1993,Schober2011,Galperin1989}.
The model assumes a bilinear coupling of the quasi-localized excitations and sound waves, which results in $\Gamma(\omega) \propto \omega^4$ 
of sound  attenuation coefficient $\Gamma(\omega)$ \cite{Schober2011,Buchenau1992}.
The strength of sound attenuation is related to the strength of the coupling and the density of the quasi-localized modes. 
Therefore, within the soft potential model, there is a link between the modes in excess of the Debye theory, the thermal conductivity plateau and sound attenuation. 


Since simulations necessarily use finite size systems, at low frequencies phonon-like modes are found in discrete locations. Thus, it is possible to
isolate and study quasi-localized modes by examining frequencies away from the phonon bands \cite{Kapteijns2018}. Using this method,
simulations have provided evidence that the density of states of quasi-localized modes scales as $\omega^4$. For three-dimensional glasses
this finding was corroborated by simulational studies that used large systems and separated modes on the basis of their participation ratio \cite{Mizuno2017,Wang2019}.
The density of states was found to be a sum of the Debye term and the density of the quasi-localized modes. Recently it was argued that this finding
is consistent with experimental results \cite{Moriel2024}.


Another approach that examines the density of states of glasses is the fluctuating elasticity theory \cite{Schirmacher2010,Schirmacher2011,Marruzzo2013,Mahajan2021}, 
which rationalizes the presence of the excess modes through frequency dependent speed of sound $c(\omega)$ and sound damping coefficient
$\Gamma(\omega)$.  Since the Debye theory predicts a relation between the density of states and the speed of sound, $g(\omega) \sim c^{-d}(\omega)$, 
a softening of the speed of sound 
with increasing frequency naturally leads to an increase in the $g(\omega)$ above the Debye theory result that is
obtained with the speed of sound that is frequency-independent.
However, comparison with simulations have shown that the softening of the speed of sound does not fully account for the excess density of states
and that one also has to include sound damping \cite{Marruzzo2013}.

Fluctuating elasticity theory expresses the density of states in terms of the Green function and provides approximate formulas for this function 
\cite{Schirmacher2006,Schirmacher2007,Marruzzo2013,Schirmacher2015}.
In an approach known as the generalized Debye formalism one postulates the form of the Green function, typically assuming damped harmonic
response of individual particles in the glass, and then uses this function in the formula for the density of states. In this way the density of
states can be related to the speed of sound and sound damping coefficient obtained from direct simulations of sound attenuation. 
The accuracy of this formalism was examined by Mizuno and Ikeda \cite{Mizuno2018}, who used a simplifying approximation
introduced by Marruzzo \textit{et al.} \cite{Marruzzo2013} in the context of the fluctuating elasticity theory analysis.
Mizuno and Ikeda used Maruzzo \textit{et al.}'s approximation and derived and studied a generalized Debye relation in two and three dimensions. 
They found that the generalized Debye relation gives an accurate description
of the boson peak height and position for their systems.  However, it is unclear 
if the agreement is within error of the simulation, if it is accurate for other systems, 
and to what extent it results from the additional approximation introduced by Maruzzo \textit{et al.}

Quasi-localized excitations assumed within the soft-potential approach, excess modes predicted by the fluctuating elasticity
theory and excess modes derived using generalized Debye relations are inter-related rather than opposing views. For example,
when one considers wave propagation in a disordered medium with varying elastic constants (which is the picture assumed within the
fluctuating elasticity approach), formulas very similar to generalized Debye relations naturally emerge \cite{Schirmacher2010},
resulting in similar predictions for the excess modes. 
However, generalized Debye relations themselves do not give any insight into  the nature of the excess modes. 

Here we investigate the following two inter-related problems. First, we use our previously derived microscopic theory of sound
attenuation \cite{Szamel2022} to identify the regions in glasses that primarily account for sound damping, which we interpret as defects. We
adopt the definition of defects that we proposed and examined in two dimensions \cite{Flenner2024} to three-dimensional glasses. In addition,
we examine two alternative definitions. We show that they mostly identify the same regions in individual glass samples.
The defects are associated with regions of large vibrational amplitude (large components of the eigenvectors of the Hessian matrix),
and hence are frequently found in the centers of a quasi-localized excitation.

Second, we examine further generalized Debye relations. We derive expressions for density of states in two and three dimensions
without the simplifying approximation introduced by Maruzzo \textit{et al.} We compare the generalized Debye predictions with
the density of states obtained directly from diagonalizing the Hessian matrix.

The paper is organized as follows. First, we describe the simulations and essential aspects of the 
microscopic sound damping theory \cite{Szamel2022}.  Next, we examine a vibrational mode level and a particle level
contribution to sound damping, which allows us to define defects. We examine three different 
definitions of defects, and find that the details change slightly but the overall interpretation is 
robust to the definition of the defects. Lastly, we examine generalized Debye relations without 
the approximation of the Green function made by previous authors, and compare our 
generalized Debye expression, a previous expression, and the density of states obtained from simulations. 
We finish with some concluding remarks.  

\section{Results and Discussion}

\subsection{Theoretical Evaluation of Sound Damping}
\label{theory}
We presented the details of the theoretical sound damping calculation in Ref. \citenum{Szamel2022}. Here we 
outline the procedure and important results. We consider the harmonic approximation with 
the initial conditions $\mathbf{u}_i(0) = \mathbf{b} e^{-i \mathbf{k} \cdot \mathbf{R}_i}$ and
$\dot{\mathbf{u}}(0) = 0$. Solving the harmonic equations of motion is equivalent to solving
$\partial_t^2 \left|1(t)\right> = -\mathbf{D}(k) \left| 1(t) \right>$, where $\mathbf{D}_{ij}(k) = \mathbf{D}_{ij} e^{i (\mathbf{R}_n - \mathbf{R}_m}$,
$\left| 1(0) \right> = N^{-1/2} (\mathbf{e}_1, ..., \mathbf{e}_N)$ and all $\mathbf{e}_n$ are $d$ dimensional 
vectors that satisfy $\mathbf{e}_n \cdot \mathbf{e}_n = 1$. 

In the small wavevector limit it can be shown \cite{Szamel2022} that sound damping is given by 
\begin{equation}
\label{smallk}
\Gamma_\lambda(\omega) = \frac{\omega^2}{v_\lambda^2} \frac{1}{\delta \omega} 
\sum_{\omega_p \in [\omega - \delta \omega/2,\omega + \delta \omega/2]} \varepsilon(\omega_p),
\end{equation}  
where,
\begin{equation}
\label{epsilon}
\varepsilon(\omega_p) = \pi/(2 \omega_p^2) \left| \left<1\right| \mathbf{X} \left| \mathcal{E}(\omega_p)\right>\right|^2
\end{equation}
and $v_\lambda$ is the speed of sound. The vector $\left| \mathcal{E} (\omega_p)\right>$ is a normalized eigenvector 
of the Hessian $\mathbf{H}$ with eigenvalue $\omega_p^2$. Recall that all masses are equal, so $\mathbf{H} = \mathbf{D}$. 
The matrix 
$\mathbf{X} = k^{-1} \mathbf{H}_{ij} \mathbf{k} \cdot (\mathbf{R}_i - \mathbf{R}_j)$ is determined by the polarization of the sound wave specified by
unit vector $\mathbf{e}_n$.
For transverse sound, $\mathbf{e}_n$ is perpendicular to $\mathbf{k}$. For example, we can set $\mathbf{e}_n = (1,0,0)$
and and then $\mathcal{X}_{ij} = \mathbf{H}_{ij}(Y_i - Y_j)$ or $ \mathbf{H}_{ij}(Z_i - Z_j)$. For longitudinal sound $\mathbf{e}_n$ is parallel to $\mathbf{k}$,
we can set  $\mathbf{e}_n = (1,0,0)$ and then $\mathbf{X}_{ij} = \mathbf{H}_{ij}(X_i - X_i)$.
The quantity $\varepsilon(\omega_p)$ is a mode level contribution to sound damping.
The definition of a mode in this work is an eigenvector of $\mathbf{H}$, which may differ from a quasi-localized excitation. Since Quasi-localized excitations can 
hybridize with plan waves, it is possible that an extended mode contains one or more quasi-localized excitations. Future work needs to examine how 
our results are related to quasi-localized excitations and not just on the structure of the eigenvectors.    

\subsection{Mode Level Contribution to Sound Damping}

\begin{figure}
\includegraphics[width=0.95\columnwidth]{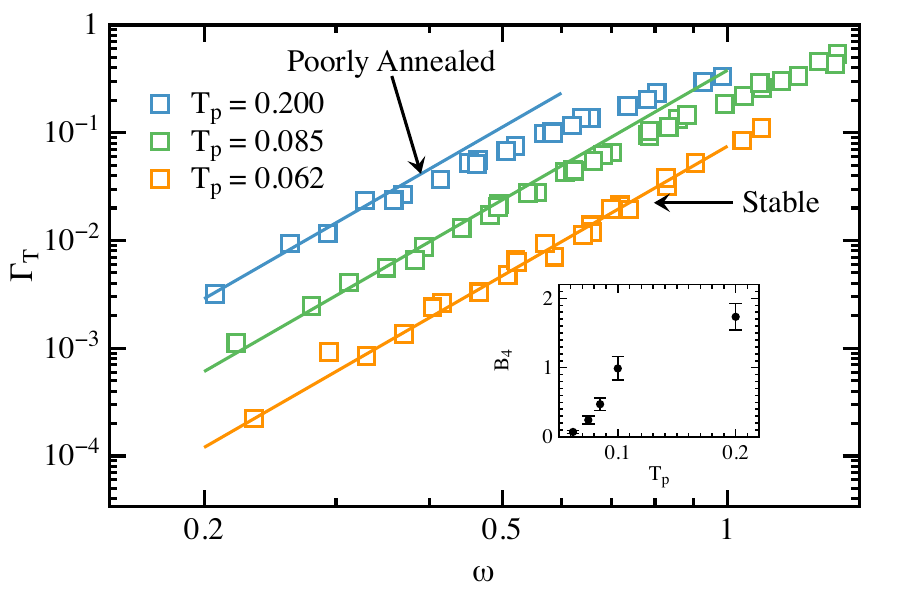}
\caption{\label{damping}Comparison of sound damping calculated from simulations, squares, to the small frequency 
theoretical predictions, lines, for three glasses of different stability. The theory predicts Rayleigh scaling at small frequencies.}
\end{figure}

Previously we have shown that the full wave-vector dependent theory accurately describes 
transverse and longitudinal sound damping in two-dimensional and three-dimensional 
glasses \cite{Szamel2022,Flenner2024}. Additionally, we found that the small wavevector approximation given in Eq.~\eqref{smallk}
accurately describes the Rayleigh scaling regime in two-dimensional glasses \cite{Flenner2024}. In practice we determine 
sound damping by evaluating the expression given in Eq.~\eqref{smallk} and fit it to 
$\Gamma(\omega) = B_4^t \omega^4$, inset to Fig.~\ref{smallk}. We note that if we allow the exponent to vary we obtain 
values very close to the expected result of 4 for Rayleigh scaling in three dimensions. Shown 
in Fig~\ref{damping} is $\Gamma$  calculated from simulations (squares) and theory (solid lines) for 
our least stable glass $T_p = 0.200$, our most stable glass $T_p = 0.062$, and a glass of 
intermediate stability $T_p = 0.085$. The agreement between the simulations and the theory is excellent
in the Rayleigh scaling regime. 

\begin{figure}
\includegraphics[width=0.95\columnwidth]{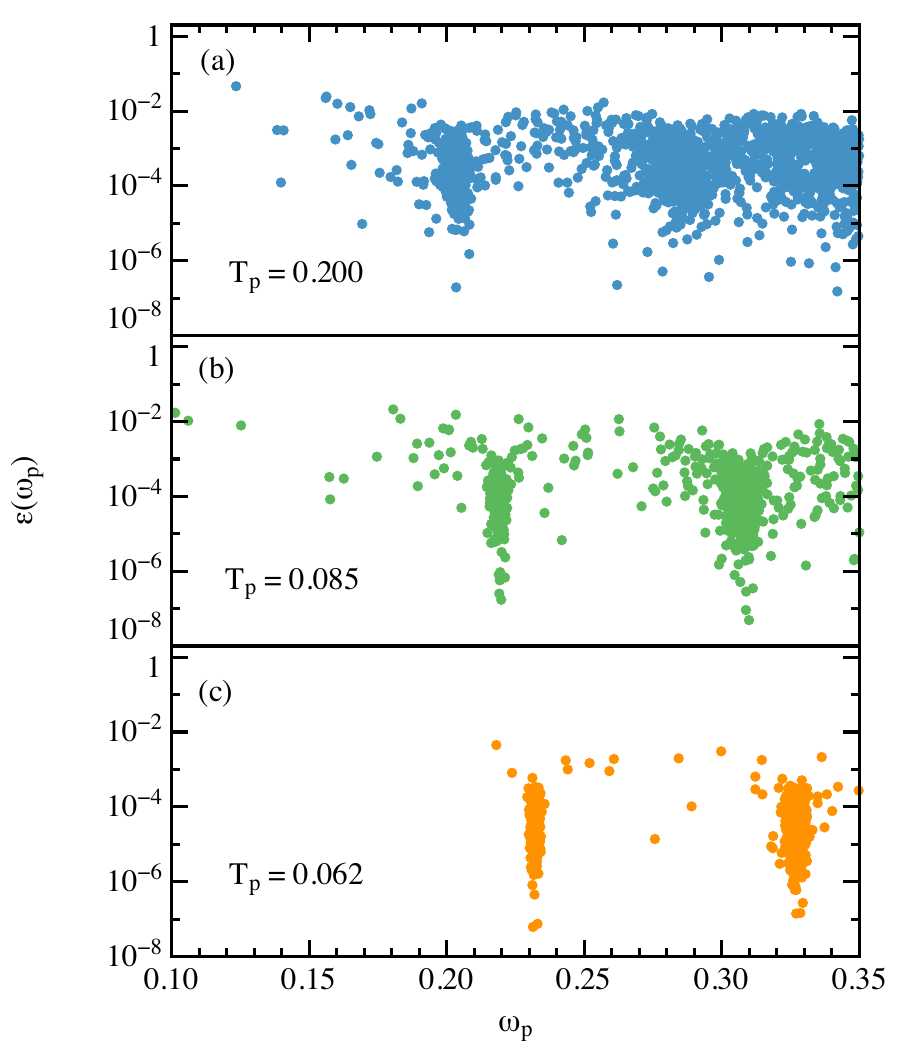}
\caption{\label{epsilonfig}Mode level contributions to sound damping for three parent temperatures of $d=3$ glasses.  The contributions
decreases with increasing stability, i.e.\ decreasing parent temperature.}
\end{figure}

To gain insight into the changes in the system which result in the large decrease in sound damping between 
our most stable and least stable glasses, we examine the mode level contribution to sound damping 
$\varepsilon(\omega_p)$, Fig.~\ref{epsilonfig}. For this calculation we are showing modes for 
the $N=192,000$ particle system where we analyzed 20 separate glass configurations for each $T_p$. 
For our most stable glass we can see bands of contributions
corresponding to modes in the first two transverse sound waves. There are also contributions from modes 
outside these bands, and these modes have a larger per mode contribution on average than the modes
within the bands. However, there are significantly fewer modes outside the phonon bands compared to the number of modes within these bands.
As the stability decreases, \textit{i.e.} with increasing $T_p$, we find that the bands become less distinguishable and
the mode level contribution increases. 

Previous work \cite{Mizuno2017,Wang2019} suggested that the low-frequency density of states can be divided into two contributions, the Debye contribution $A_D \omega^2$
and an excess contribution $A_4 \omega^4$, such that the full density of states is $g(\omega) = A_D \omega^2 + A_4 \omega^4$.
In glasses \cite{Mizuno2017,Wang2019,Kapteijns2018},
there are contributions to sound damping arising from the Debye-like modes and from the excess modes. We studied these
contributions in two-dimensional glasses and showed that non-affine forces play a key role in understanding the contribution from Debye-like modes, 
and there is sound damping even if the eigenvectors in expression \eqref{epsilon}
were replaced with sound waves\cite{Flenner2024}. 
It has been argued by several authors that non-affine forces play a key role in sound damping\cite{Caroli2020,Baggioli2022}
We were also able to identify regions of the glass
that were responsible for large sound damping over a range of frequencies, which we referred to as sound damping defects. To define a defect
we defined a particle level contribution to sound damping for each mode, and determined that the defects were associated with regions
where the particles average square magnitude of the displacement over a frequency range was greater than $2/N$, \textit{i.e.}\ the maximum displacement for an 
ideal plane wave. Here we use this approach in three-dimensional glasses to reveal regions of individual configurations that give large contributions to sound damping.

The mode level contribution is proportional to the projection of the non-affine force field on the eigenvectors of the Hessian, 
$\left|\mathcal{E}(\omega_p)\right>$. For a transverse wave, which is shown here, it is the non-affine force
field due to a shear deformation. For a longitudinal wave, it is the non-affine force field due to a compression. 
We can see that the frequency dependence arises from the change of the eigenvectors with frequency. Here we 
will examine the contributions to transverse waves since they are the lowest frequency waves. 

In Fig.~\ref{epsilonfig}(c), $T_p = 0.062$, there is one eigenvector with a larger contribution than the other eigenvectors. 
For this $T_p$ we can count the number of modes up to $\omega = 0.24$, right past the first phonon band. 
We find that there is one more mode than expected from the Debye theory, but we do not know which mode 
to identify as the excess mode. We determined which glass configuration contained this mode, and examined the contribution 
to sound damping for the first 13 modes, which should encompass the first phonon band plus one extra mode. 

\begin{figure}
\includegraphics[width=0.95\columnwidth]{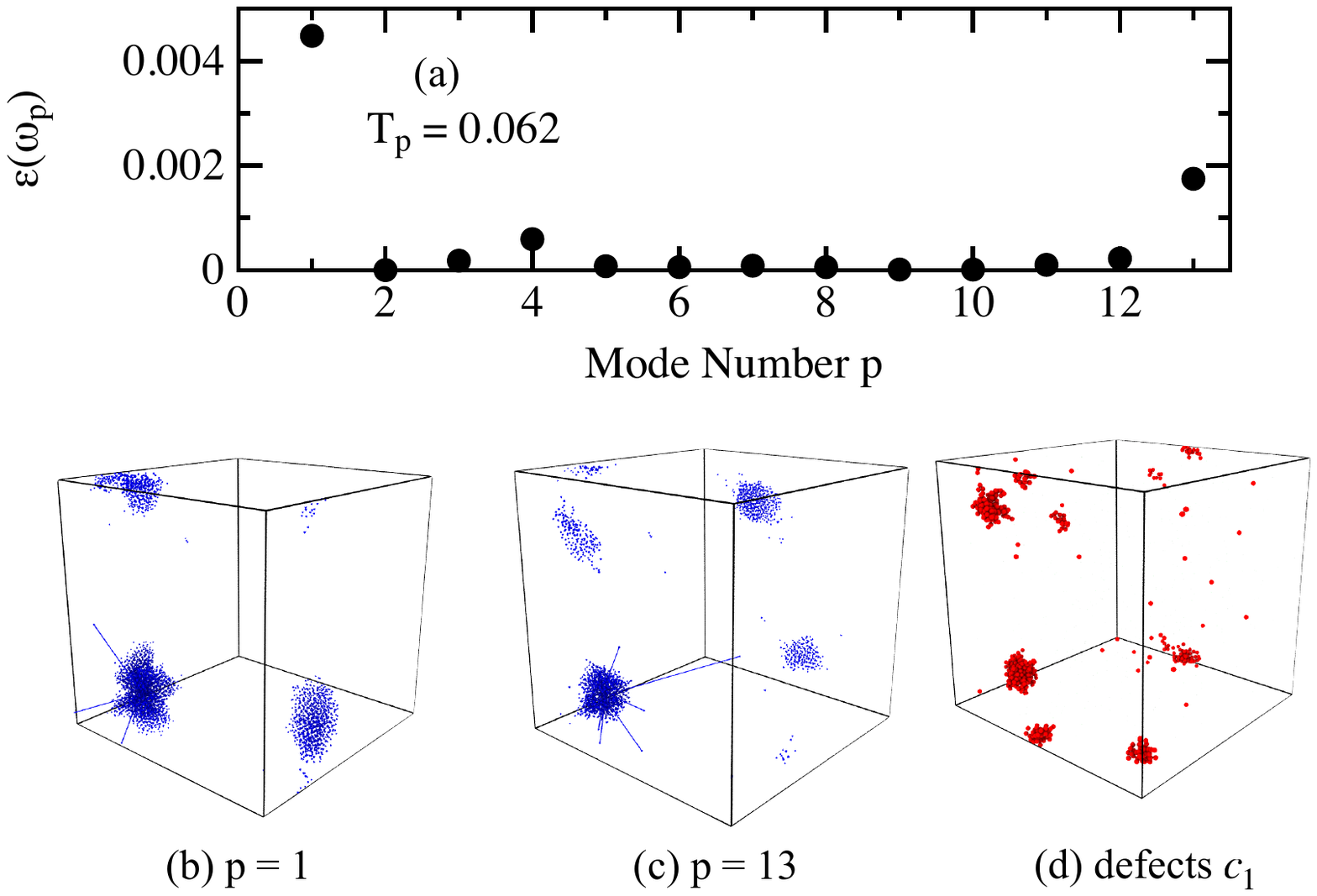}
\caption{\label{vis1}(a) The mode level contributions to sound damping for one $T_p = 0.062$ configuration. The modes are ordered from
the lowest frequency to the highest frequency mode, excluding the uniform translations. (b) Visualization of the lowest frequency mode, $p=1$,
which has the largest contribution to sound damping. Only the displacement vectors with magnitude $|\mathbf{E}_n|^2 > 2/N$ are shown. 
(c) Visualization of the mode $p=13$, and the same region of space as for $p=1$ has large eigenvectors. (d) The defects
identified using our first procedure, see text, and we see that this procedure identifies the compact area of large displacement 
magnitude shown in (b) and (c) as a defect.}
\end{figure}

Shown in Fig.~\ref{vis1}(a) is the contribution to sound damping for this one $T_p = 0.062$ glass configuration versus the mode number,
where the modes are numbered from lowest frequency to highest frequency (excluding the uniform translations). We see that the first
mode has the largest contribution to sound damping, and the 13th mode also has a relatively large contribution. To examine the structure
of the mode we only look at the displacement vectors of particle $n$, $|\mathbf{E}_n|^2 > 2/N$, i.e.\ the $|\mathbf{E}_n|^2$ that are larger than for a 
pure plane wave. We note that we observe regions with large $|\mathbf{E}_n|^2$ in mode 13 different from what is seen in mode 1. 

Shown in Fig.~\ref{vis1}(b) is the lowest frequency mode that has the largest contribution to sound damping for the $T_p = 0.062$ glass configuration 
shown in Fig.~\ref{vis1}(a). There is a cluster of particles with larger $|\mathbf{E}_n|^2$, which resembles a quasi-localized excitation. We note that
the participation ratio for this mode is $10^{-4}$, and the mode is more localized than the other modes. We also find a cluster of 
large $|\mathbf{E}_n|^2$ in mode 13 at the same location as in mode 1 for the same glass configuration. As we found in two-dimensional glasses, one
region of space can result in large $|\mathbf{E}_n|^2$ over a range of frequencies. This observation motivates the definition of defects we 
discuss in the next section. 

\begin{figure}
\includegraphics[width=0.95\columnwidth]{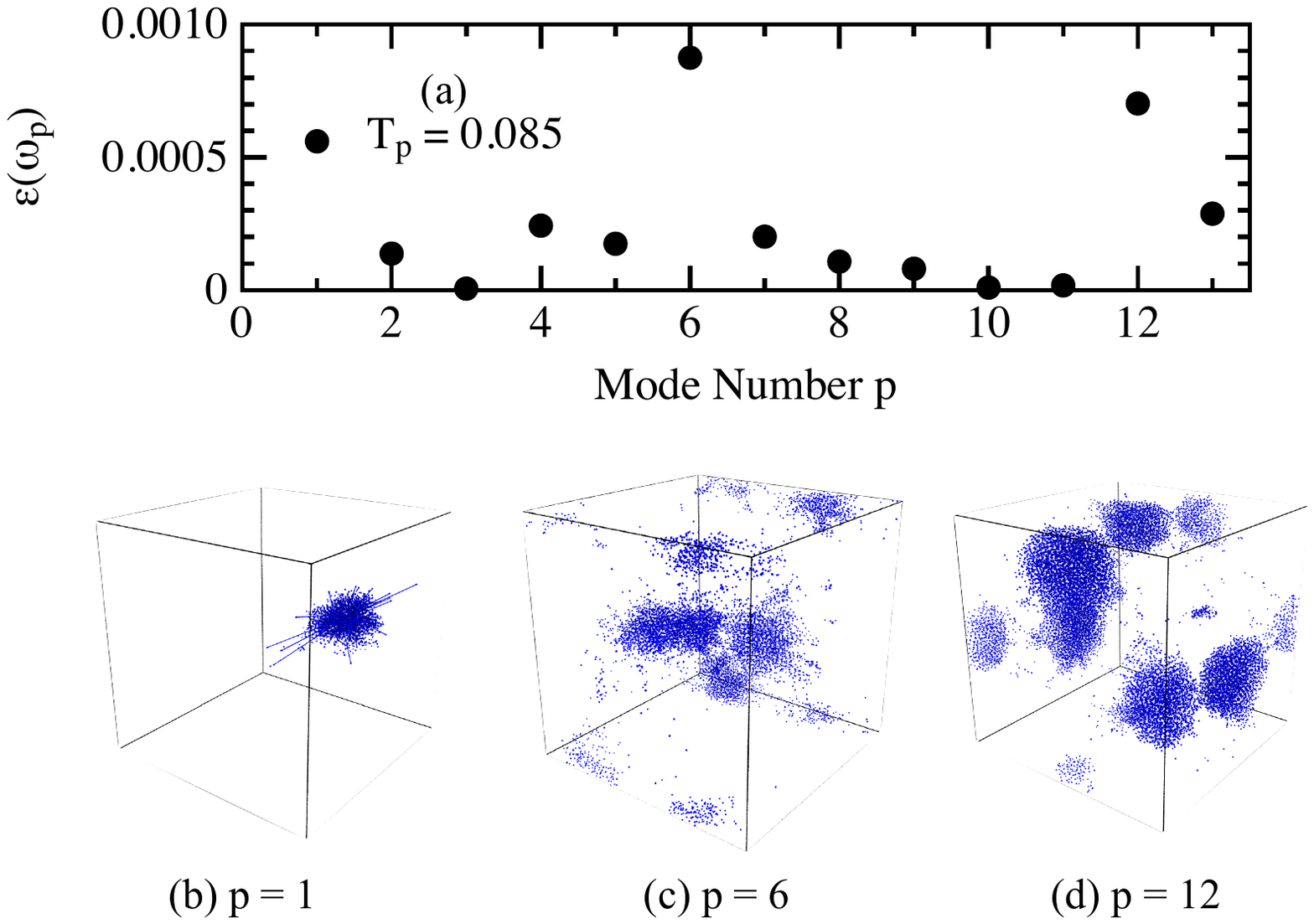}
\caption{\label{vis2}(a) The mode level contributions to sound damping for one $T_p = 0.085$ configuration. The modes are ordered from
the lowest frequency to the highest frequency mode, excluding the uniform translations. (b) The mode with the third largest contribution to 
sound damping, $p=1$, which is a compact area that resembles a quasi-localized excitation. Only the displacement vectors with magnitude
$|\mathbf{E}_n|^2 > 2/N$ are shown. (c) Mode with the largest largest contribution to 
sound damping for this configuration, $p=6$. There are no compact areas of large displacements, and the eigenvectors where  $|\mathbf{E}_n|^2 > 2/N$
are not localized. (d) Mode with the second largest contribution to 
sound damping for this configuration, $p=6$. There are no compact areas of large displacements, and the eigenvectors where  $|\mathbf{E}_n|^2 > 2/N$
are not localized. More extended modes can have large contributions to sound damping. The defect particles are determined through an average 
particles with $|\mathbf{E}_n|^2$ over a range of frequencies, and this configuration has several defects.}
\end{figure}

While we find that, in general, eigenvectors that have a large contribution to sound damping have a cluster of particles with large $|\mathbf{E}_n|^2$,
we also find that there are also configurations where 
the largest contribution to sound damping comes from a mode where there is no compact cluster of large $|\mathbf{E}_n(\omega_p)|^2$. One example
is shown in Fig.~\ref{vis2}. The contribution to sound damping is shown in Fig.~\ref{vis2}(a) for one configuration for a parent temperature $T_p = 0.085$. 
The mode that makes the largest contribution is $p=6$, the next largest is $p=12$, and the third largest is $p=1$. 
The first mode, Fig.~\ref{vis2}(b) (p=1), has one compact cluster of large $|\mathbf{E}_n|^2$, while although the other two modes show regions of space with 
large $|\mathbf{E}_n|^2$, these regions are not compact and isolated. We note that the participation ratio is $10^{-4}$ for $p=1$, 0.61 for $p = 6$, and 0.64 for $p=12$. 
Thus the participation ratio for $p=6$ and $p=12$ is what is expected for a plane wave, while the participation ratio is what would be expected for a 
quasi-localized mode for $p=1$.  We note, however, that a hybridized quasi-localized excitation can still be present in modes $p=6$ and $p=12$.

While we can associate eigenvectors with large $|\mathbf{E}_n|^2$ with eigenvectors that also have a large contribution to sound damping,
it is not clear if these eigenvectors are distorted plane waves or contain quasi-localized excitations. 
Just by observation, we find that modes with small participation ratio have a large contribution 
to sound damping, but the reverse is not true. Modes with participation ratios close to what is expected for a plane wave can also have 
large contributions to sound damping. Future work needs to examine definitions of quasi-localized excitations and particle level contributions 
to sound damping. 

\subsection{Defects} 

\begin{figure}
\includegraphics[width=0.95\columnwidth]{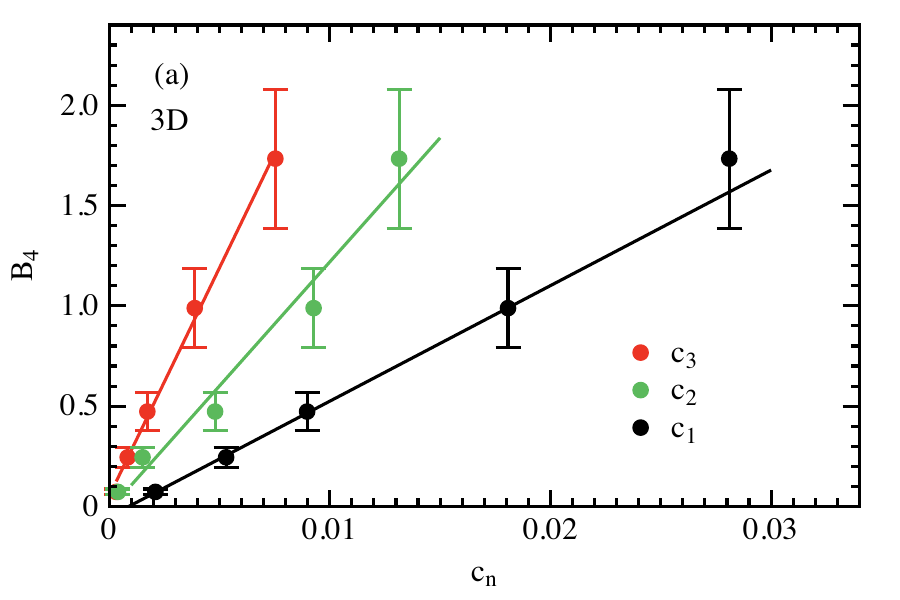}
\includegraphics[width=0.95\columnwidth]{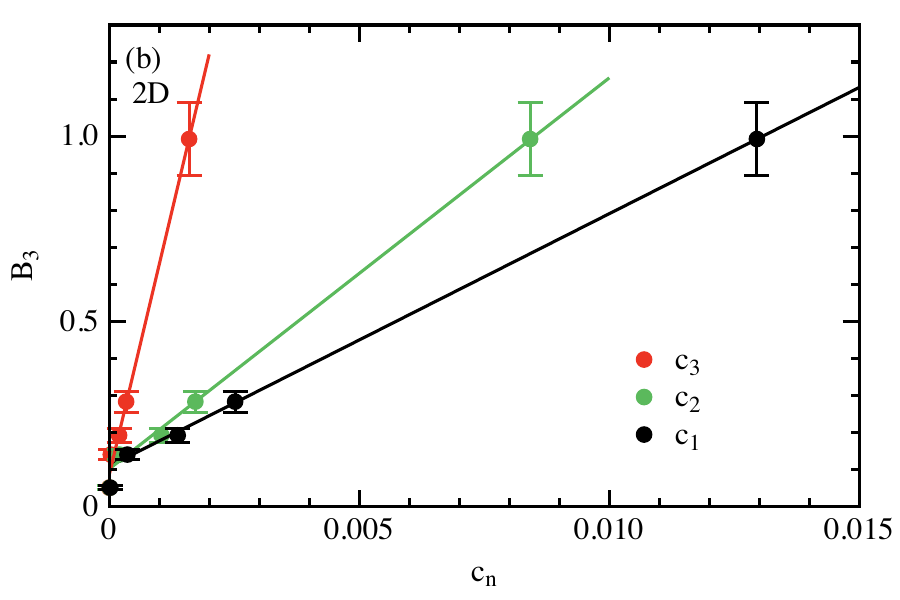}
\caption{\label{defects}The Rayleigh scaling coefficient versus the 
fraction of the particles within a defect with the three definitions of defects described in the text in three dimensions (a) and two dimensions
(b). For the three-dimensional glasses, the Rayleigh scaling coefficient goes to zero when the defect density goes to zero, which suggests that 
defects control sound damping for these three dimensional glasses. For our two-dimensional glasses the Rayleigh scaling coefficient is non-zero
even when the defect density goes to zero. Defects are not needed for sound damping to occur, and we expect this to be the case for
some three-dimensional glasses also.}
\end{figure}

For two-dimensional amorphous
solids we previously defined a particle $n$ to be within a defect if $S_n^{2d} = (N/2) \sum_{p=1}^{24} |\mathbf{E}_n(\omega_p)|^2/24 > 1$,
where where $\mathbf{E}_n(\omega_p)$ is the component of the eigenvector with eigenvalue $\omega_p^2$ corresponding to particle $n$ \cite{Flenner2024}.
This definition closely resembles the vibriality defined in by Hu and Tanaka\cite{Hu2022} and studied by Richard\cite{Richard2020}, and is related to the region where sound attenuation begins 
as identified by Mahajan and Pica Ciamarra\cite{Mahajan2024}.
The frequency range 
was chosen to coincide with the range of the Rayleigh scattering regime for each $T_p$ and to include the 
same number of modes in the analysis for each configuration. We note that the Rayleigh scattering regime 
extends to higher $\omega$ than what is included in the analysis for our most stable glasses. We then defined 
a defect density as $c_1 = \bar{w}_n /N$ where $w_n$ is one for a defect particle and zero otherwise. The over-bar denotes an
average over glass configurations. We examined the relationship between $c_1$ and fits to $\Gamma = B_3 \omega^3$ from
$\Gamma$ determined from simulations. We found that the 
most stable glasses in two-dimensions had no defects, Fig.~\ref{defects}(b) (black circles) using this definition. Additionally, the
defect concentration $c_1$ is linearly related to $B_3$. 

We acknowledge that there are some arbitrary choices in the definition of defects. To examine how robust our conclusions are on the 
definition of defect particles through the size of $|\mathbf{E}_n(\omega_p)|^2$ and the range of frequencies used to define defects, we examine two additional definitions of
defects for the two-dimensional and three dimensional solids. The corresponding defect densities are denoted by $c_2$ and $c_3$. 

We start with defects in three-dimensional glasses defined in a very similar manner as in our previous work in 2D and define 
$S_n^{3d} = (N/2) \sum_{p=1}^{52} |\mathbf{E}_n(\omega_p)|^2/52 > 1$ for $N = 192,000$. 
Again, the range was chosen to 
roughly correspond to the Rayleigh scaling regime for the three-dimensional glasses and to include the same 
number of modes for each glass. If there are no excess modes, the first 52 modes are the first 4 transverse sound waves.
The longitudinal waves for a finite sized systems show up at higher frequencies and are not included in our analysis. 
We had to choose this large system to average over the Rayleigh scaling regime since Rayleigh scaling occurs only at
small frequencies for the poorly annealed glasses. We defined $c_1 = \bar{w}_n/N$ where $w_n$ is one when $S_n^{3D}$ is 
greater than one and zero otherwise.  
We examine the concentration of defects $c_1$ to fits of $\Gamma = B_4 \omega^4$ using the simulation data.
We find that $c_1$ is linearly related to $B_4$, we also find that a linear fits implies no damping even 
with defect particles, black circles and line in Fig.~\ref{defects}(a).
While this conclusion is different than what we concluded in 2D, we note
that there are several technical issues that may arise with this definition. 

First, it is somewhat ambiguous what range of frequencies we should use to define defects. To test how this effects the results we 
defined a defect where the average is taken over the first 8 modes in two dimensions (roughly the first two transverse phonon bands),
and the first 30 modes in three-dimensions (roughly the first two transverse phonon bands). We defined $c_2$ using this restricted 
region of modes and found that the linear relationship still held in two-dimensions and three dimensions, green symbols and line in Fig.~\ref{defects}. 
Here the two-dimensional results still predict damping with no defects, and the three-dimensional result predicts that damping goes to zero when the defects
concentration goes to zero. 

We also examined a more restrictive definition of a defect. We found the particles where the average of $|\mathbf{E}_n(\omega_p)|^2$ was 
greater than $2/N$ for the first phonon band, and separately for the second phonon band. The defect particles was the intersection of these two
sets of particles and the defect concentration is denoted as 
$c_3$, red symbols in Fig.~\ref{defects}. Again we find damping without defects in two dimensions, but damping approximately goes to zero 
with the defect concentration in three dimensions. 

\begin{figure}
\includegraphics[width=0.95\columnwidth]{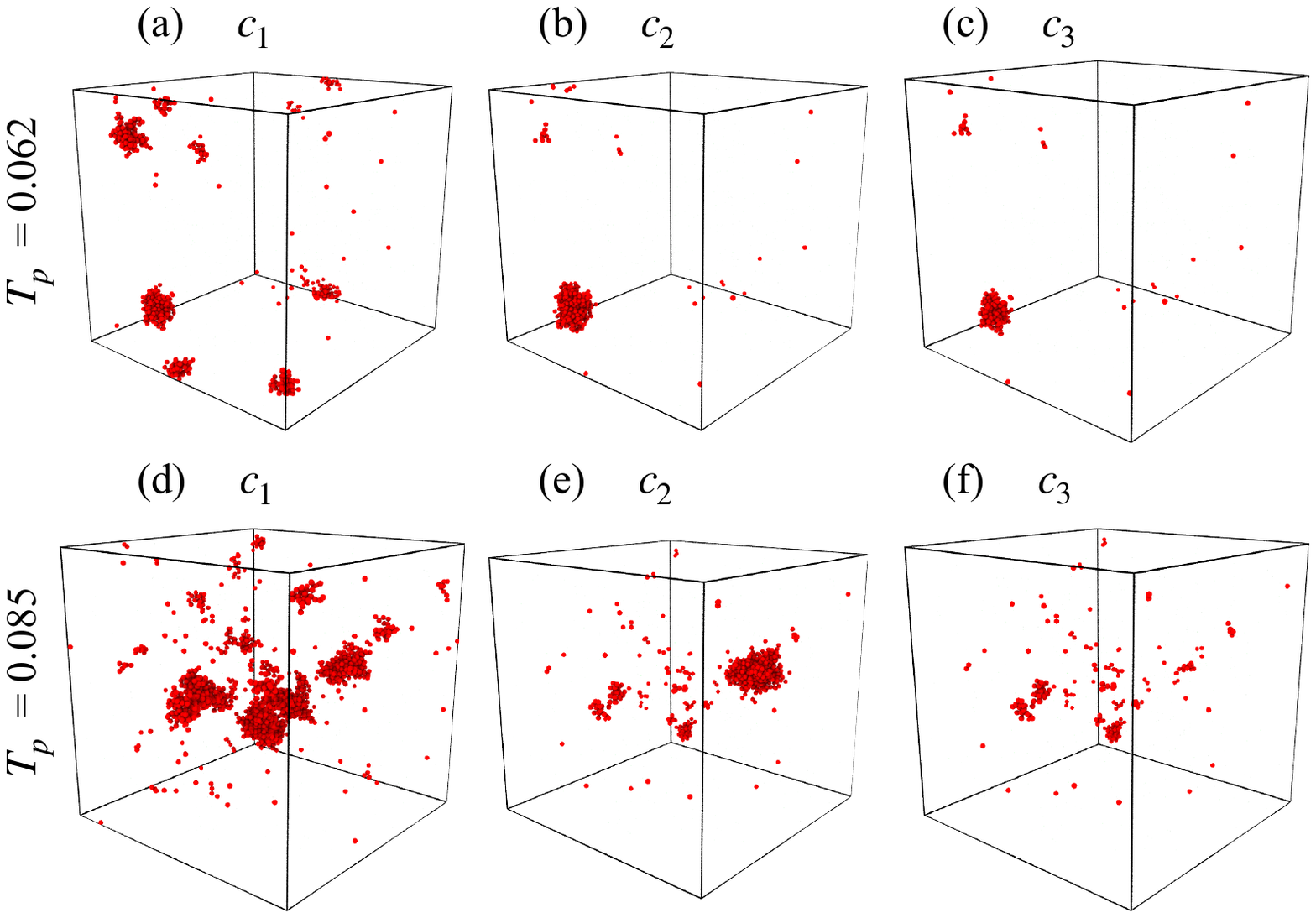}
\caption{\label{visdefect} Visual representation of defects for one configuration for our three definitions of defects and two 
parent temperatures. The configurations are the same shown in Fig.~\ref{vis1} and Fig.~\ref{vis2}. (a) $T_p = 0.062$ and 
defect definition $c_1$. (b) $T_p = 0.062$ and defect definition $c_2$. (c) $T_p = 0.062$ and 
defect definition $c_3$. (d) $T_p = 0.085$ and defect definition $c_1$. (e) $T_p = 0.085$ and defect definition $c_2$. (f) $T_p = 0.085$ and defect definition $c_3$.}
\end{figure}

To get a better understanding of the different definitions of defects, we visualize the particles within the defects for $T_p = 0.062$ 
and $T_p = 0.085$, Fig.~\ref{visdefect}. We choose the same configurations  shown in Fig.~\ref{vis1} and Fig.~\ref{vis2}. We find that 
as the definition of the defect becomes more strict, with $c_1$ being the least restrictive and $c_3$ being the most restrictive,
the fraction of particles found in a defect decreases. For $T_p = 0.062$ there are defects that are picked up by including more than the first two
phonon bands in the definition of a defect, $c_1$, that are not present in the other two definitions. For all definitions we do find a few isolated
particles, which are small contributions to the total fraction of defects. We also show the defects determined from definition $c_1$ in 
Fig.~\ref{vis1}(d), where we can see a correspondence of the areas with large $|\mathbf{E}_n|^2$ and the location of defects.  

We conclude that defects play a large role in sound damping for our simulated three-dimensional solid. However, as in two dimensions, there is no
reason to believe that defects have to be present for sound damping to occur. Future work studying different systems and densities in three 
dimensions should help clarify this issue. We note, however, that large, stable systems are needed to properly define defects in the
Rayleigh scattering regime. Another possibility would be to artificially reduce quasi-localized modes through the reduction of internal stresses\cite{Lerner2018,Kapteijns2021}, 
verify that these modes are indeed related to the sound damping defects, and then study damping with defect density.  
 
 \subsection{Sound Damping and the Density of States}

 Within a generalized Debye model, as presented in Sec. 2.2 of Ref. \citenum{SchirmacherScopignoRuocco2015}, density of states can be obtained from
 frequency dependent speeds of sound and sound damping coefficients. 
 \begin{equation}
 \label{gint}
 g(\omega) = \frac{2 \omega}{\pi q_D^d}\int_0^{q_D} dq q^{d-1} \mathrm{Im}[(d-1)G_T(q,\omega) + G_L(q,\omega)],
 \end{equation}
 where the Green function reads
 \begin{equation}
 \label{gf}
 G_\lambda(q,\omega) = \frac{1}{-\omega^2 + q^2 v_\lambda^2(\omega)(1-i \Gamma_\lambda(\omega)/\omega)},
 \end{equation}
 with $\lambda$ denoting the transverse, $\lambda\equiv \mathrm{T}$ and longitudinal, $\lambda\equiv \mathrm{L}$ components.  Furthermore,
 $q_D$ in Eq. \eqref{gint} denotes the Debye wavevector, $q_D = (2 d \pi^{d-1} \rho)^{1/d}$.
 %
 We will follow this somewhat phenomenological approach here. In our comparison used frequency dependent speeds
 of sound and sound damping coefficients determined from sound damping simulations \cite{Wang2019a}.
 
 We note that to simplify the integral over wavevectors, previous authors\cite{Marruzzo2013,Mizuno2018} approximated the denominator in Eq. \eqref{gf}  
 $-\omega^2+q^2v^2(1-i\Gamma/\omega) \approx (1-i\Gamma/\omega)(-\omega^2+q^2v^2)$.
 To study this approximation as well as the generalized Debye model in more detail we 
 did not make this approximation and evaluated Eq.~\eqref{gint} with Eq.~\eqref{gf} as the integrand.

 Since the formulas for the density of states are rather complicated, to simplify the notation we omitted the frequency dependence of the
 speeds of sound and sound attenuation coefficients. 
 
\begin{figure}
 \includegraphics[width=0.95\columnwidth]{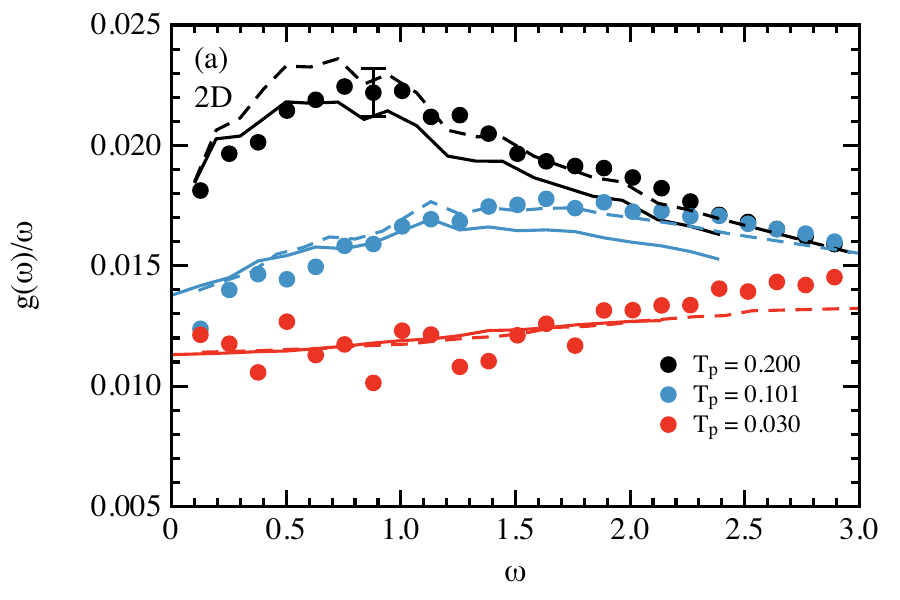}
 \includegraphics[width=0.95\columnwidth]{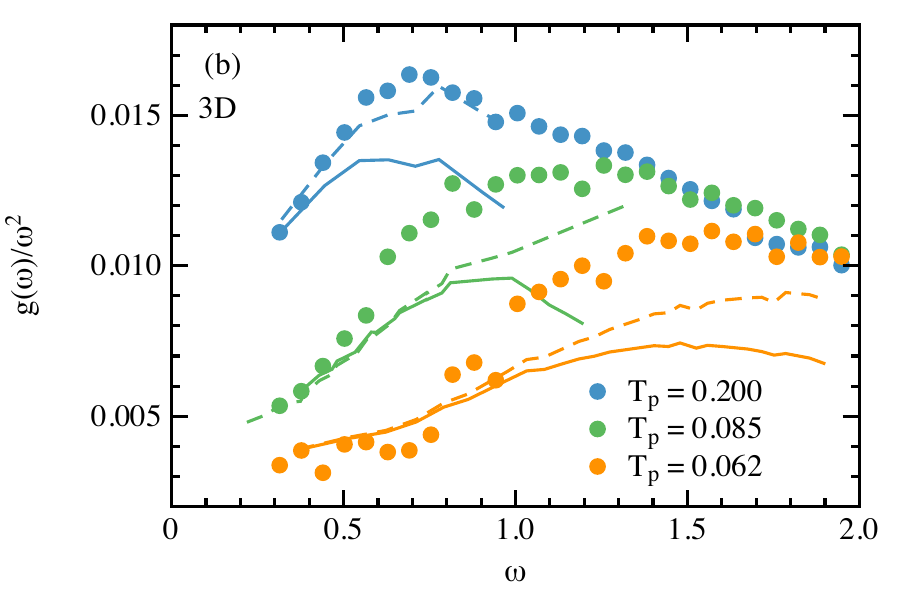}
 \caption{\label{gd}Comparison of the reduced density of states (circles) with generalized Debye relations (lines) for three-dimensional (a)
 and two-dimensional (b) glasses. The solid lines is the generalized Debye relation derived in this work, while the dashed line is the generalized
 Debye relation given by Mizuno and Ikeda.}
 \end{figure}

 In two-dimensions we obtained the following result, $g^{2D}(\omega) = g_T^{2D}(\omega) + g_L^{2D}(\omega)$ where
 \begin{eqnarray}
 \label{gd2D}
 g_\lambda^{2D}(\omega) &=& \left( \frac{\omega}{q_D^2 v_\lambda^2 (1 + \Gamma_\lambda^2/\omega^2)} \right)
 + \frac{\omega}{\pi q_D^2 v_\lambda^2 (1+\Gamma_\lambda^2/\omega^2)} \tan^{-1}\left(\frac{ \Gamma_\lambda q_D^2 v_\lambda^2}{\omega (\omega^2 - q_D^2 v_\lambda^2)}\right) 
 \nonumber \\
 && + \frac{\Gamma}{2 \pi q_D^2 v_\lambda^2 (1+\Gamma_\lambda^2/\omega_\lambda^2)} \log\left[ 1 - \frac{ 2 v_\lambda^2 q_D^2}{\omega^2} 
 + \frac{v_\lambda^4 q_D^4}{\omega^4} \left(1+ \frac{\Gamma_\lambda^2}{\omega^2} \right) \right],
 \end{eqnarray} 
 and we recall that $\Gamma_\lambda,$ and $v_\lambda$ depend on frequency $\omega$.
 When $\Gamma_\lambda = 0$ the Debye theory $g(\omega) = [(q_D^2 v_T^2)^{-1} + (q_D^2 v_L^2)^{-1}] \omega$ is recovered. 
 When $\Gamma_\lambda > 0$ the Debye term is reduced by a factor of $(1+\Gamma_\lambda^2/\omega^2)$,
 and is also influenced by the frequency change of $v_\lambda(\omega)$. For many glasses there is a softening of the speed of sound resulting in a 
 reduction of  $v_\lambda(\omega)$, which would increase the density of states.  
 Both these contributions are negligible at small frequencies. There are two additional terms that result in an increase of the density of states. 
 The $\tan^{-1}$ term is small and the $\log$ term is nearly equal to the $\log$ term given by Mizuno and Ikeda \cite{Mizuno2018}. Therefore, this expression gives 
 very similar results to the expression in Eq.~(15) of Mizuno and Ikeda \cite{Mizuno2018}. 
 We compare Eq.~\eqref{gd2D} with the 
 two-dimensional density of states in Fig.~\ref{gd}(a), and find excellent agreement with the density of states calculated directly from simulations. 
 We note that Mizuno and Ikeda's expression (dashed line) is just as accurate as our expression.
 This results suggests 
 that sound damping accounts for the boson peak, at least for two-dimensional glasses.
 
 However, we find a different result for three-dimensional glasses. For three-dimensional glasses the generalized Debye model leads to 
 $g^{3D}(\omega) = 2g_T^{3D}(\omega) + g_L^{3D}(\omega)$ where the rather complex expressions for $g_\lambda^{3D}(\omega)$
 are presented in the appendix. 
 
 There are two simplifications that we can make to understand the result given in the appendix. First, we note that in the absence of sound damping the
 Debye theory is recovered, $g_\lambda^{3D}(\omega) = [2(q_D^3 v_T^3)^{-1} + (q_D^3 v_L^3)^{-1}]\omega^2$.
 Again, with sound damping there is a competition between an increase in $\Gamma_\lambda(\omega)$ and a decrease in $v_\lambda(\omega)$.  Second, if 
 $\Gamma_\lambda << \omega$ then $A_\lambda \approx 1$, $B_\lambda \approx 0$ and we have the simple result
 \begin{equation}
 g_\lambda^{3D}(\omega) \approx \frac{\omega^2}{q_D^3 v_\lambda^3} + \frac{2 \Gamma_\lambda}{\pi q_D^2 v_\lambda^2 (1+\Gamma_\lambda^2/\omega^2)},
 \end{equation}
 which is the formula given by Maruzzo \textit{et al.} \cite{Marruzzo2013}. 
The modes in excess of the Debye theory are proportional to the sound damping coefficients and would scale as $\omega^4$ in three dimensions. If we 
approximate $1 + \Gamma_\lambda^2/\omega^2$ with 1 then we find that the excess density of states scales as sound damping.

When we approximate $1 + \Gamma_\lambda^2/\omega^2$ with 1 and consider low frequencies, we are left with a term linear in $\omega$
and a term that grows as $\Gamma(\omega) \sim \omega^3$. 
This would 
suggest that the excess density of states for two-dimensional glasses scale differently than the excess density of states for
three-dimensional glasses \cite{Flenner2024,Wang2022,Wang2021}. Additionally, this does not agree with the $\omega^4$ scaling 
of the density of states observed by Kapteijns \textit{et al.}\cite{Kapteijns2018} 

Unlike in two dimensions, the generalized Debye relation does not reproduce the boson peak for
our three dimensional glasses, Fig.~\ref{gd}(b). While the low frequency data is well produced,
the generalized Debye relation underestimates the boson peak height. We note that the expression for $g(\omega)$ given by Mizuno and Ikeda,
dashed lines in Fig.~\ref{gd}(b),  results in predictions for the boson peak that are larger than
our result without this approximation. For our least stable glass, $T_p = 0.200$, the boson peak is well reproduced using Mizuno and Ikeda's formula, but 
the boson peak is still underestimated for the other parent temperatures.

It is currently unclear if the success of the generalized Debye relation is related to any fundamental difference in two dimensions and three dimensions, or if 
this difference is just coincidental. Mizuno and Ikeda found that their generalized Debye relation reproduced the boson peak for two-dimensional and three-dimensional
glasses. However, their approximation underestimated the density of states at small frequencies for their three-dimensional glasses. We find that 
our expression accurately predicts the small frequency density of states for each of our glasses. The accuracy of the generalized Debye model and the range of 
frequencies where it would be expected to hold deserves further study. 

\section{Discussion}

We previously proposed a definition of defects in two-dimensional glasses and showed that sound is damped and Rayleigh scaling obeyed
even if no defects are present \cite{Flenner2024}. Here we extended defect definition to three-dimensional glasses and, in addition,
proposed two additional, more stringent definitions. We showed that these definitions identify defects mostly 
in the same locations in individual glass configurations. The spatial extend of these defects decreases with increasing stringency of the definition.

We note that unlike for two-dimensional glasses, defects are always present in our three-dimensional glasses. This may be due to the
fact that one can reach lower parent temperatures and thus more stable glasses in two dimensions compared to three dimensions.
Thus, to check whether sound is damped and Rayleigh scaling obeyed in three-dimensional glasses without defents one may
resort to removing artificially internal stresses, following Kapteijns \textit{et al.}\cite{Lerner2018,Kapteijns2021}.

We also examined the ability of generalized Debye relations to predict the boson peak. We derived a 
generalized Debye formula without resorting to an approximation made by previous authors.
Our result for the excess density of states is smaller than previous result that followed from an approximate evaluation of the integral.
In two-dimensions the change is minimal, but it is significant in three dimensions where we found 
that our expression begin to noticeably deviate from previous expressions when $\Gamma/\omega > 0.1$.
Additionally, we found that all the expressions under-estimated the boson peak height for our stable 
three-dimensional glasses. We note that the expression given by Mizuno and Ikeda accurately predicts
the boson peak height for our least stable three-dimensional glass. 

The formulation of generalized Debye models relies on assumptions of the wavevector dependence of the 
Green function (or response function).  Mizuno and Ikeda \cite{Mizuno2018} replaced the 
frequency of the damped sound wave with $v_\lambda(q,\omega) q$ in one place in their formulation
of the Green function to introduce the wavevector dependence of the 
Green function. Additionally, they dropped the $q$ dependence of the speed of 
sound. While this assumption is reasonable at low frequencies, it likely breaks down upon approaching 
the boson peak. It is 
possible that a more accurate formulation of the Green function
for transverse waves around the boson peak 
will allow generalized Debye models to be more accurate. Another possibility is the existence of additional 
vibrational modes that are not captured in the generalized Debye models. 

The nature of the vibrational modes around the
boson peak have been studied by Hu and Tanaka in two-dimensional glasses\cite{Hu2022} and three-dimensional glasses\cite{Hu2023},
where they claim that stringlet like excitations are responsible for the boson peak. In contrast, Lerner and Bouchbinder claim that 
the boson peak is composed of hybridized phononic and quasilocalized excitations for two-dimensional glasses\cite{Lerner2023}. Future work needs to 
examine the connection between sound attenuation, stringlet like excitations, and quasilocalized excitations. 

\section*{Acknowledgements}

We gratefully acknowledge the support of NSF Grant No. CHE 2154241.

\section{Appendix}

\subsection{Simulations}
Our glasses were created from equilibrated configurations of a polydisperse glass forming liquid with particles of equal mass \cite{Ninarello2017}. The interaction potential is
given by
\begin{equation}
\label{pot}
V(r_{ij}) = \left( \frac{\sigma_{ij}}{r_{ij}}\right)^{12} + c_0 + c_2 \left( \frac{r_{ij}}{\sigma_{ij}} \right)^2 + c_4 \left( \frac{r_{ij}}{\sigma_{ij}} \right)^{4},
\end{equation}
when $r_{ij} \le 1.25 \sigma_{ij}$ and zero otherwise. 
In Eq.~\eqref{pot} $r_{ij} = |\mathbf{r}_i - \mathbf{r}_j|$ is the distance between particles 
$i$ and $j$. The parameters $c_0$, $c_2$, and $c_4$ are chosen such that the potential and its first two derivatives are continuous at $1.25 \sigma_{ij}$.
The probability of a particle diameter $\sigma$ is $P(\sigma) \sim \sigma^{-3}$ where $\sigma \in [0.73,1.63]$ and we use the
non-additive mixing rule $\sigma_{ij} = 0.5(\sigma_i + \sigma_j)(1-0.2|\sigma_i - \sigma_j|)$. The number density is $N/V \equiv \rho = 1$.
We studied systems of sizes $N$ = 192,000,  64,000, and 48,000 particles in three dimensions and $N$ = 20,000 particles in two dimensions.
The Swap Monte Carlo algorithm \cite{Ninarello2017} was used to create equilibrated configurations at a parent temperature $T_P$.
These configurations were then quenched to zero temperature using FIRE algorithm \cite{Bitzek2006} implemented in LAMMPS \cite{lammps}.
For reference, in three dimensions the onset temperature of 
slow dynamics is $T_o = 0.200$, the mode-coupling temperature is $T_c = 0.108$, and the estimated experimental glass temperature is 
$T_g = 0.072$ \cite{Ninarello2017}. In two dimensions $T_o = 0.200$, $T_c = 0.183$, and $T_g = 0.082$ \cite{Berthier2019}. 

To determine sound attenuation in the harmonic approximation we followed the procedure proposed by Gelin \textit{et al.} \cite{GelinNatMat2016}.
We numerically solved the harmonic equations of  motion 
\begin{equation}
\ddot{\mathbf{u}_i}(t) = -\sum_{j=1}^N \mathbf{D}_{ij} \cdot \mathbf{u}_j(t) + \dot{u}_i^0 \delta(t),
\end{equation}
where $\mathbf{D}_{ij}$ is the dynamical matrix (equal to the Hessian since all masses are equal) and
$\mathbf{u}_i(t)$ denotes the displacement of the particle $i$ at $t$ from its inherent structure position \cite{Mizuno2018,Wang2019a}. 
The initial condition $\dot{\mathbf{u}}_i^0 = \mathbf{a}_\lambda(\mathbf{k}\cdot\mathbf{r}_i^0)$
is an excited sound wave at $t = 0$. Transverse sound waves are simulated when $\mathbf{a}_T \cdot \mathbf{k} = 0$
and longitudinal sound waves are studied when $\mathbf{a}_L \propto \mathbf{k}$. Sound damping $\Gamma$ 
and the frequency of sound $\Omega_\lambda$ are
calculated from fitting the velocity correlation function 
\begin{equation}
v_\lambda(t) = \frac{
\sum_{i=1}^N \dot{\mathbf{u}}_i(0) \cdot \dot{\mathbf{u}}(t)}
{\sum_{i=1}^N \dot{\mathbf{u}}_i(0) \cdot \dot{\mathbf{u}}(0)}
\end{equation}
to $v_\lambda(t) = \exp(-\Gamma_\lambda t/2) \cos(\Omega_\lambda t)$. 
In practice we use an envelope method described by Wang \textit{et al.}~\cite{Wang2019a} to find 
$\Gamma_\lambda$ and then fix $\Gamma_\lambda$ in the fits to find $\Omega_\lambda$. 

\subsection{Generalized Debye Derivation}

Generalized Debye relations determine the density of states using 
\begin{equation}
\label{dosint}
g(\omega) = \frac{2 \omega}{\pi q_D^d} \int_0^{q_D} \mathrm{Im}\left[ (d-1) G_T(q,\omega) + G_L(q,\omega)\right] q^{d-1} dq,
\end{equation}
where $G_\lambda(q,\omega)$ is the Green function for phonon propagation. The Green function can be determined from several
different starting points. Mizuno and Ikeda started by defining the Green function in the time domain as the 
integral of $C(t)$ and assuming $C(t) = e^{-\Gamma t/2} \cos(\Omega t)$. Under some assumptions they arrive at the expression
\begin{equation}
G_\lambda(q,\omega) = \frac{1}{-\omega^2 + q^2v_\lambda^2(\omega)(1-i\Gamma_\lambda/\omega)}. 
\end{equation}
Fluctuating elasticity theory derives an equivalent expression \cite{Schirmacher2006,Schirmacher2007,Schirmacher2015,Marruzzo2013}, 
but their starting point is the wave equation assuming 
fluctuating elastic constants. Previous work made the approximation 
$-\omega^2 + q^2 v_\lambda^2(\omega)(1-i\Gamma_\lambda/\omega) \approx (1-i\Gamma_\lambda/\omega)(-\omega^2 + q^2 v_\lambda^2)$,
which should hold for $\Gamma_\lambda << v_\lambda$. Using this approximation for the denominator, the integral \eqref{dosint} is easier to 
evaluate. 

To evaluate integral \eqref{dosint} without further approximations
we make the change of variables $x = -\omega^2 + q^2v_\lambda^2$ and take the imaginary part after performing the 
integral. This procedure is rather straightforward in two dimensions, but is involved and leads to a complex equation in three dimensions. The solution in
three dimensions is
 \begin{eqnarray}
 g_\lambda^{3D}(\omega) &=& \frac{\omega^2}{\pi q_D^3 v_\lambda^3 (1 + \Gamma_\lambda^2/\omega^2)}
 \left[(A_\lambda^3 - 3A_\lambda B_\lambda^2) \theta + (B_\lambda^3 - 3 A_\lambda^2 B_\lambda) \ln(r_\lambda) \right] \nonumber \\
 &&+\frac{2 \Gamma_\lambda}{\pi q_D^2 v_\lambda^2(1+\Gamma_\lambda^2/\omega^2)} \\
 A_\lambda & = & \sqrt{\frac{\sqrt{1+\Gamma_\lambda^2/\omega^2}+1}{2}} \\
 B_\lambda & = &  \sqrt{\frac{\sqrt{1+\Gamma_\lambda^2/\omega^2}-1}{2}} \\
 \theta_\lambda &=& \pi + \tan^{-1}\left(\frac{q_D v_\lambda \omega \sqrt{2}\sqrt{\sqrt{1+\Gamma_\lambda^2/\omega^2}-1}}{\omega^2 - q_D^2 v_\lambda^2 \sqrt{1+\Gamma_\lambda^2/\omega^2}} \right) \\
 r_\lambda &=& \frac{\sqrt{\omega^4 - 2 \omega^2q_D^2v_\lambda^2 + (1+\Gamma_\lambda^2/\omega^2) q_D^4 v_\lambda^4}}{\omega^2 - 2 q_D v_\lambda \omega A_\lambda +q_D^2 v_\lambda^2 \sqrt{1+\Gamma_\lambda^2/\omega^2}}.
 \end{eqnarray}
The Debye term is recovered without damping since $A_\lambda = 1$ , $B_\lambda = 0$, and the $\tan^{-1}$ term is zero. As shown in Fig.~\ref{gd}, the small $\omega$ 
solution given here and by Mizuno and Ikeda\cite{Mizuno2018} are nearly identical. We find that our result starts to deviate from 
that of Mizuno and Ikeda when $\Gamma/\omega > 0.1$ for our three-dimensional glasses. 
Marruzzo \textit{et al.}\ \cite{Marruzzo2013} gave a slightly different expression than Mizuno and Ikeda, but all our expressions are nearly equal at small $\omega$.



\begin{thebibliography}{99}
\bibitem{Ramos2023}
M. A. Ramos (ed.), \textit{Low-Temperature Thermal and Vibrational Properties of Disordered Solids}, World Scientific Publishing Europe Ltd., London (2023).
\bibitem{Zeller1971}
R.C. Zeller and R.O. Pohl, Thermal conductivity and specific heat of noncrystalline solids, Phys. Rev. B \textbf{4}, 2029-2041 (1971).
\bibitem{Buchenau1984}
U. Buchenau, N. N\"ucker, and A.J. Dianoux, Neutron Scattering Study of the Low-Frequency Vibrations in Vitreous Silica, Phys. Rev. Lett. \textbf{53}, 2316-2319 (1984).
\bibitem{Schroeder2004}
J. Schroeder, W. Wu, J.L. Apkarian, M. Lee, L.G. Hwa, and C.T. Moynihan, Raman scattering and Boson peaks in glasses: temperature and pressure effects,
J. Non-Cryst. Solids \textbf{349}, 88-97 (2004).
\bibitem{Moriel2024} 
A. Moriel, E. Lerner, and E. Bouchbinder, Boson peak in the vibrational spectra of glasses, Phys. Rev. Research \textbf{6}, 023053 (2024).
\bibitem{Karpov1982}
V.G. Karpov, M.I. Klinger, and F.N. Ignatiev, Atomic tunneling states and low-temperature anomalies of thermal properties in amorphous materials, Solid State Commun. \textbf{44}, 333 (1982)
\bibitem{Buchenau1992}
U. Buchenau, Yu. M. Galperin, V.L. Gurevich, D.A. Parshin, M.A. Ramos, and H.R. Schober, Interaction of soft modes and sound waves in glasses, Phys. Rev. B \textbf{46}, 2798 (1992).
\bibitem{Schober2011}
H.R. Schober, Quasi-localized vibrations and phonon damping in glasses, J. Non-Cryst. Solids \textbf{357}, 501-505 (2011).
\bibitem{Galperin1989}
Yu. M. Galperin, V.G. Karpov, and V.I. Kozub, Localized states in glasses, Adv. in Phys. \textbf{38}, 669 (1989).
\bibitem{Ramos1993}
M.A. Ramos, L. Gil, A. Bringer, and U. Buchenau, The density of tunneling and vibrational states of glasses within the soft-potential model, Phys. Stat. Sol.(a) \textbf{135}, 477 (1993).
\bibitem{Kapteijns2018}
G. Kapteijns, E. Bouchbinder, and E. Lerner, Universal Nonphononic Density of States in 2D, 3D, and 4D Glasses, Phys. Rev. Lett. \textbf{121}, 055501 (2018). 
\bibitem{Mizuno2017}
H. Mizuno, H. Shiba, and A. Ikeda, Continuum limit of the vibrational properties of amorphous solids, Proc. Natl. Acad. Sci. \textbf{114}, E9767 (2017).
\bibitem{Wang2019}
L. Wang, A. Ninarello, P. Guan, L. Berthier, G. Szamel, and E. Flenner, Low-frequency vibrational modes of stable glasses, Nat. Commun. \textbf{10}, 26-33 (2019).
\bibitem{Schirmacher2010}
W. Schirmacher, C. Tomaras, B. Schmid, G. Baldi, G. Viliani, G. Ruocco, and T. Scapigno, 
Sound attenuation and anharmonic damping in solids with correlated disorder, Condens. Matter Phys. \textbf{13}, 23606 (2010).
\bibitem{Schirmacher2011}
W. Schirmacher, Some comments on fluctuating-elasticity and local oscillator models for anomalous vibrational excitations in glasses, J. Non-Cryst. Solids \textbf{357}, 518-523 (2011).
\bibitem{Marruzzo2013}
A. Marruzzo, W. Schirmacher, A. Fratalocchi, and G. Ruocco, Heterogeneous shear elasticity of glasses: the origin of the boson peak, Scientific Reports \textbf{3}, 1407 (2013).
\bibitem{Mahajan2021}
S. Mahajan and M.P. Ciamarra, Unifying Description of the Vibrational Anomalies of Amorphous Materials, Phys. Rev. Lett. \textbf{127}, 215504 (2021).
\bibitem{Schirmacher2015}
W. Schirmacher, T. Scopigno, and G. Ruocco, Theory of vibrational anomalies in glasses, J. Non-Cryst. Solids \textbf{407}, 133 (2015).
\bibitem{Schirmacher2006}
W. Schirmacher, Thermal conductivity of glassy materials and the "boson peak", Europhys. Lett. \textbf{73}, 892 (2006).
\bibitem{Schirmacher2007}
W. Schirmacher, G. Ruocco, and T. Scopigno, Acoustic Attenuation in Glasses and its Relation with the Boson Peak, Phys. Rev. Lett. \textbf{98}, 025501 (2007).
\bibitem{Mizuno2018}
  H. Mizuno and A. Ikeda, Phonon transport and vibrational excitations in amorphous solids, Phys. Rev. E \textbf{98}, 062612 (2018).
  \bibitem{Szamel2022}
  G. Szamel and E. Flenner, Microscopic analysis of sound damping in low-temperature amorphous solids reveals quantitative importance of non-affine effects, 
  J. Chem. Phys. \textbf{156}, 144502 (2022).
\bibitem{Flenner2024}
E. Flenner and G. Szamel, The Origin of Sound Damping: Defects and Beyond, arXiv:2406.18667 (2024). 
\bibitem{Caroli2020}
C. Caroli and A. Lemaitre, Key role of retardation and non-locality in sound propagation in amorphous solids as evidenced by a projection formalism, J. Chem. Phys. \textbf{153}, 144502 (2020).
\bibitem{Baggioli2022}
M. Baggioli and A. Zaccone, Theory of sound damping in amorphous solids from nonaffine motions, J. of Phys.: Condens. Matter \textbf{34}, 215401 (2022).
\bibitem{Hu2022}
Y.C. Hu and H. Tanaka, Origin of the boson peak in amorphous solids, Nature Physics \textbf{18}, 669-677 (2022).
\bibitem{Richard2020}
D. Richard, M. Ozawa, S. Patinet, E. Stanifer, B. Shang, S.A. Ridout, B. Xu, G. Zhang, P.K. Morse, J.-L. Barrat, L. Berthier, M.L. Falk, P. Guan, A.J. Liu, K. Martens, S. Sasty, 
D. Vandembroucq, E. Lerner, and M.L. Manning, Predicting plasticity in disordered solids from structural indicators, Phys. Rev. Mater. \textbf{4}, 113609 (2020). 
\bibitem{Mahajan2024}
S. Mahajan and M. Pica Ciamarra, Heterogeneous attenuation of sound waves in three-dimensional amorphous solids, Phys. Rev. E \textbf{109}, 024605 (2024). 
  \bibitem{Ninarello2017}
A. Ninarello, L. Berthier, and D. Coslovich, Models and algorithms for the next generation of glass transition studies, Phys. Rev. X \textbf{7}, 021039 (2017).
 \bibitem{Bitzek2006}
E. Bitzek, P. Koaskinen, F. G\"ahler, M. Moseler, and P. Gumbsch, Structural relaxation made simple, Phys. Rev. Lett. \textbf{97}, 170201 (2006).
\bibitem{lammps}
A.P. Thompson, H.M. Aktulga, R. Berger, D.S. Bolintineanu, W.M. Brown, P.S. Crozier, P.J. in't Veld, A. Kohlmeyer, S.G. Moore, T.D. Nguyen, R. Shan, M.J. Stevens, J. Tranchida, 
C. Trott, S.J. Plimpton, LAMMPS - a flexible simulation tool for particle-based materials modeling at the atomic, meso, and continuum scales, Comp. Phys. Comm. \textbf{271}, 10817 (2022).
\bibitem{Berthier2019}
  L. Berthier, P. Charbonneau, A. Ninarello, M. Ozawa, and S. Yaida, Zero-temperature glass transition in two dimensions, Nat.Commun. \textbf{10}, 1508 (2019).
  \bibitem{GelinNatMat2016} S. Gelin, H. Tanaka and A. Lema\^{i}tre,
Anomalous phonon scattering and elastic correlations in amorphous solids,
Nature Materials \textbf{15}, 1177 (2016).
\bibitem{Wang2019a}
L. Wang, L. Berthier, E. Flenner, P. Guan, and G. Szamel, Sound attenuation in stable glasses, Soft Matter \textbf{15}, 7018-7025 (2019).
\bibitem{Lerner2018}
E. Lerner and E. Bouchbinder, Frustration-induced internal stresses are responsible for quasilocalized modes in structural glasses, Phys. Rev. E \textbf{97}, 032140 (2018).
\bibitem{Kapteijns2021}
G. Kapteijns, D. Richard, E. Bouchbinder, and E. Lerner, Elastic moduli fluctuations predict wave attenuation rates in glasses, J. Chem. Phys. \textbf{154}, 081101 (2021). 
\bibitem{SchirmacherScopignoRuocco2015} W. Schirmacher, T. Scopigno, and G. Ruocco,
Theory of vibrational anomalies in glasses, J. Non-Cryst. Solids \textbf{407}, 133 (2015).
\bibitem{Wang2021}
L. Wang, E. Flenner, and G. Szamel, Low-Frequency Excess Vibrational Modes in Two-Dimensional Glasses, Phys. Rev. Lett. \textbf{127}, 248001 (2021).
\bibitem{Wang2022}
L. Wang, E. Flenner, and G. Szamel, Scaling of the non-phononic spectrum of two-dimensional glasses, J. Chem. Phys. \textbf{158}, 126101 (2023).
\bibitem{Hu2023}
Y.C. Hu and H. Tanaka, Universality of stringlet excitations as the origin of the boson peak of glasses with isotropic interactions, Phys. Rev. Research \textbf{5}, 023055 (2023).
\bibitem{Lerner2023}
E. Lerner and E. Bouchbinder, Boson-peak vibrational modes in glasses feature hybridized phononic and quasilocalized excitations, J. Chem. Phys. \textbf{158}, 194503 (2023).
\end{thebibliography}

\end{document}